\documentstyle[11pt,newpasp,epsf,twoside]{article}
\def\la{\mbox{\raisebox{-0.1ex}{$\scriptscriptstyle \stackrel{<}{\sim}$\,}}}

\def\avcn{\mbox{$ { \rm \overline { C_n^2 } } $ }}

\newcommand{\cn}{\mbox{ ${\rm C_n^2}$ }}

\newcommand{\nd}{\mbox{$ \nu _d $ }}

\newcommand{\ndmeas}{\mbox{$ \nu _{d,meas} $}}
\newcommand{\ndpred}{\mbox{$ \nu _{d,pred} $}}

\begin{document}

\title{Pulsar Scintillation Studies and Structure of the Local Interstellar Medium} 

\author{N. D. Ramesh Bhat}
\affil{Max-Planck-Institut f\"ur Radioastronomie, Bonn, Germany}

\author{Yashwant Gupta, A. Pramesh Rao}
\affil{National Centre for Radio Astrophysics, Pune, India}

\author{P. B. Preethi}
\affil{Hindustan College of Engineering, Madras, India}

\begin{abstract}
Results from new observations of pulsars using the Ooty Radio Telescope (ORT) are used for investigating
the structure of the Local Interstellar Medium (LISM).
The observations show anomalous scintillation effects towards several nearby pulsars, and these are 
modeled in terms of large-scale spatial inhomogeneities in the distribution of plasma density fluctuations 
in the LISM.
A 3-component model, where the Solar neighbourhood is surrounded by a shell of enhanced plasma turbulence,
is proposed for the LISM. The inferred scattering structure is strikingly similar to the Local Bubble.
Further, analysis based on recent scintillation measurements show evidence for enhanced scattering 
towards pulsars located in the general direction of the Loop I Superbubble. The model for the LISM
has been further extended by incorporating the scattering due to turbulent plasma associated with 
Loop I.
\end{abstract}

\begin{figure}[t]
\plotfiddle{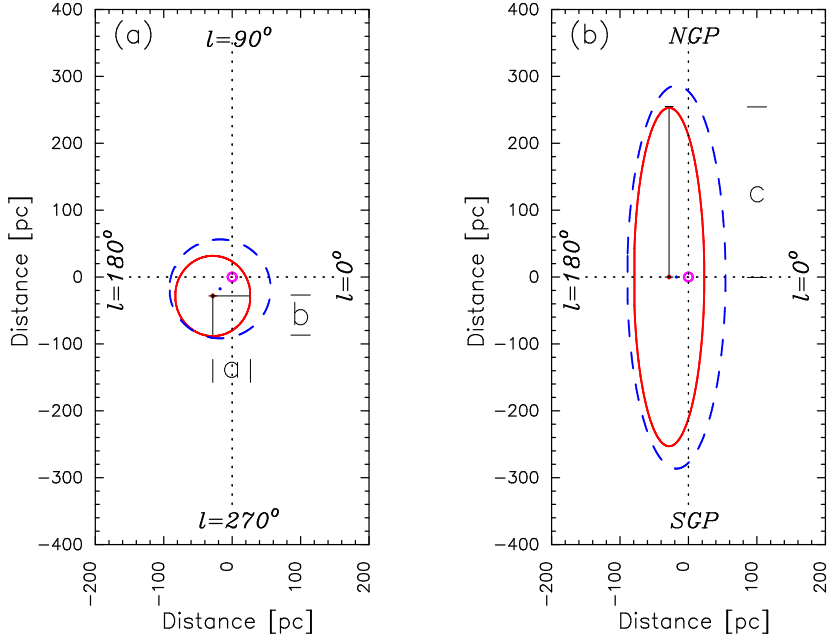}{3.25in}{270}{65}{65}{-215}{280}
\caption[]{Geometry of the local scattering structure which best reproduces the observational results.
Panels (a) and (b) are sections along the Galactic plane and along a plane perpendicular
to the Galactic plane and passing through the Galactic poles, respectively.}
\end{figure}

\section{Introduction}

Propagation effects on radio signals from pulsars, such as dispersion and scattering,
are very useful in probing the distribution of thermal plasma in the ISM.
Studies of Interstellar Scattering (ISS) can be used to understand the distribution 
and the spectrum of plasma density fluctuations.
Observational data from a variety of wavebands (ranging from X-ray to optical) suggest that the 
Solar system resides in a low-density, X-ray emitting cavity of size $\sim$ a few 100 pc 
(Cox \& Reynolds 1987; Snowden et al. 1990).
It is reasonable to expect the Local Bubble and its environment to play a substantial role in
determining the scintillation properties of nearby (\la 1 kpc) pulsars.
Such pulsars, therefore, form potential tools for studying the structure and properties of the LISM.

Scintillation properties of pulsars are studied using their dynamic scintillation spectra
$-$ records of intensity variations in the frequency-time plane.
Such spectra display intensity patterns that fade over short time intervals and narrow frequency ranges.
The average characteristics of scintillation patterns are quantified using the two-dimensional (2D) auto 
co-variance function (ACF).
The ACF is fitted with a 2D elliptical Gaussian, to yield the parameters, $viz.,$
decorrelation bandwidth ($ \nu _d $), scintillation timescale ($\tau _d$)
and the drift slope ($ d t / d \nu $).
From the present observations, it has been possible to estimate the scintillation properties and ISM
parameters with accuracies much better than that which has been possible from most earlier data.


\begin{figure}[t]
\plotfiddle{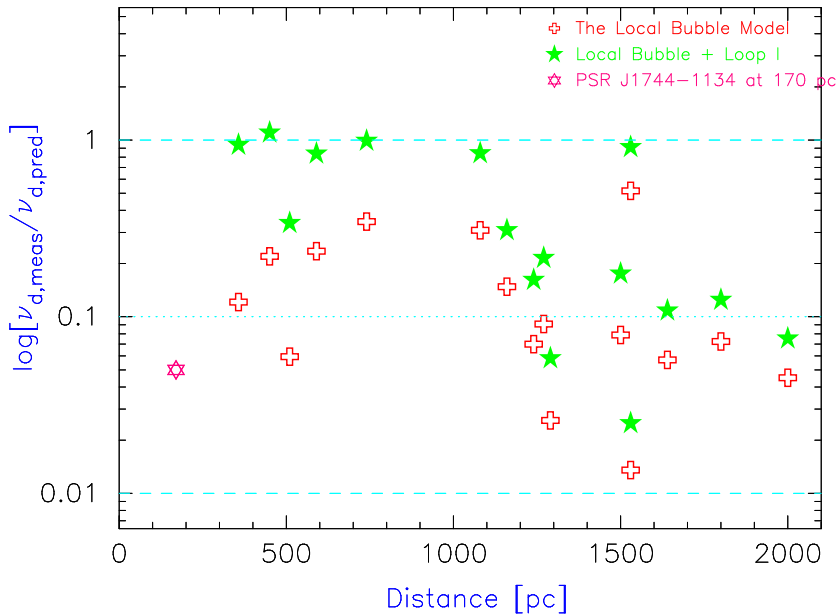}{3.25in}{270}{50}{50}{-215}{270}
\caption[]{Plots illustrating the enhanced scattering of nearby pulsars due to the Local Bubble and 
the Loop I Bubble. Ratios of the measured decorrelation bandwidths (\ndmeas) to their values as 
predicted by the models for the structure of the Local ISM (\ndpred) are plotted against the 
distance estimates. The measurements of \nd are scaled to 327 MHz.}
\end{figure}

\section{Modeling the Structure of the Local ISM}

The behaviour of the line-of-sight$-$averaged strength of scattering, \avcn, with direction ($l,b$) and
location
$-$ dispersion measure (DM) or distance (D) $-$ forms powerful means of investigating the nature of distribution
of
plasma
density fluctuations in the ISM (Cordes et al. 1991).
From our measurements of $ \nu _d $,
we have obtained very precise estimates of \avcn.
There is about two orders of magnitude fluctuations of \avcn,
which is much larger than that predicted by the current models for the \cn
distribution in the Galaxy (Cordes et al. 1991).
In addition, there is a systematic variation of \avcn with distance, whereby there is a turnover near $\sim$  200
pc,
followed by a downward trend up to $\sim$ 1 kpc.
Our analysis also shows that many of the nearby pulsars are more strongly scattered than expected.
Further, there are several cases of pulsars with comparable DMs and/or at similar distances showing 
remarkably different scintillation properties.
The strength of such anomalous scattering ($ A_{dm} $) shows a systematic behaviour with DM and distance.
These results suggest that the distribution of scattering material in the LISM is non-uniform,
and it is likely to be in the form of a large-scale coherent density structure.

The results are modeled in terms of large-scale spatial inhomogeneities in the distribution
of radio wave scattering material in the LISM.
To explain the observations, we need a 3-component model, where the Solar system resides in
a weakly scattering medium, surrounded by a shell of enhanced scattering, embedded in the normal,
large-scale ISM (Bhat et al. 1998).
The best fit values of the parameters of the local scattering structure are such that the observed trends
of $ A_{dm} $ and $ \nu _d $ are reproduced.
It has an ellipsoidal morphology, and is more extended away from the Galactic plane (Figure 1).
The centre has an offset $\sim$ 20$-$35 pc from the Sun, towards $ 215^o < l < 240^o $ and $ -20^o < b < 20^o $.
The strength of plasma density fluctuations in the shell material
($10^{-0.96}~<~ {\huge \int} _0^d~C_n^2(z)~dz~<~10^{-0.55}~{\rm pc~m^{-20/3}}$, where $d$ is the thickness of the
shell) is much larger than that in the interior ($10^{-4.70}~<~ \overline { C_n^2 } ~<10^{-4.22} ~{\rm m^{-20/3}}$)
and that in the ambient medium ($ \overline { C_n^2 } ~<~10^{-3.30}~{\rm m^{-20/3}}$).
A detailed comparison study with other pertinent studies of the LISM shows
that the morphology of the inferred scattering structure is very similar to that of the Local
Bubble, as known from observational data at X-ray, EUV, UV and optical bands.

\section{Further Extensions of the Local ISM Model}

Observational data in the recent past have considerably improved our understanding of the structure of the 
LISM (cf. Breitschwerdt et al. 1998). 
In particular, there has been clear evidence for interaction between the Local Bubble and nearby Loop I 
Superbubble (Egger 1998). The Loop I Bubble, with a size $\sim$ 300 pc and located at $\sim$ 170 pc
towards $l \sim 330^o$, $b \sim 20^o$ (Sco-Cen OB association), covers a large region of the sky (angular diameter
$\sim 120^o$), and hence likely to play a substantial role in the dispersion and scintillation of nearby pulsars.
In order to examine this, we carried out a critical comparison study of recent scintillation measurements for
pulsars whose lines of sight intersect the Loop I Bubble with their predictions from the Local Bubble model. 
The scintillation data used in our analysis are from Johnston et al. (1998), Bhat et al. (1999) and Cordes (1986).
Our analysis reveal large discrepancies between the measurements and predictions, whereby the level of enhanced 
scattering is much higher than that suggested by the Local Bubble model (Figure 2).

The proposed LISM model is further extended by modeling the distribution of turbulent plasma associated with the
Loop I Bubble. Scintillation data of pulsars PSR J1744$-$1134 and PSR J1456$-$6843 (at parallax distances of 357 pc 
and 450 pc, respectively) are used to constrain the strength of scattering in the Loop I boundary. 
PSR J1744$-$1134 is a very interesting object in this context. 
Toscano et al. (1999) report precise estimates of parallax and proper motion for this pulsar.
With the new distance estimate ($ 357 \pm 35 $ pc), the second boundary of Loop I is located 
at mid-way ($ \approx $ 175 pc) towards this pulsar, suggesting that the enhanced level of scattering reported for 
this pulsar (\nd $\approx$ 300 kHz, cf. Johnston et al. 1998) is 
predominantly due to the Loop I shell.
Such enhanced scattering could not have been explained if the pulsar was to be located at 170 pc
(indicated by {\it unfilled star} in Figure 2), 
as suggested by the Taylor \& Cordes (1993) model, whereby the pulsar has to be located in the interior of Loop I.
The inferred strength of scattering in the Loop I shell is found to be somewhat larger than that in the Local 
Bubble shell. 
The improved LISM model can successfully explain the enhanced scattering of most pulsars within $\sim$ 1 kpc 
(Figure 2).


\end{document}